\documentclass[twocolumn,showpacs,amsmath,amssymb]{revtex4}
\usepackage{mathrsfs}
\usepackage{amsmath}

\usepackage{graphicx}
\usepackage{amsfonts}% Include figure files
\usepackage{dcolumn}% Align table columns on decimal point
\usepackage{bm}% bold math

\begin{document}

\title{How does degree heterogeneity affect nucleation of Ising model on complex networks?}

\author{Hanshuang Chen$^{1}$} %\email{chenhshf@mail.ustc.edu.cn}

\author{Shuxian Li$^{2}$}

\author{Gang He$^1$}

\author{Feng Huang$^3$}

\author{Chuansheng Shen$^4$}

\author{Zhonghuai Hou$^2$}\email{hzhlj@ustc.edu.cn}

\affiliation{$^{1}$School of Physics and Material Science, Anhui
University, Hefei 230039, People's Republic of China \\
$^2$Hefei National Laboratory for Physical Sciences at
 Microscales \& Department of Chemical Physics, University of
 Science and Technology of China, Hefei 230026, People's Republic of China \\
 $^3$Department of Mathematics and Physics, Anhui University of Architecture, Hefei 230601, People's Republic of China \\
 $^4$Department of Physics, Anqing Teachers College, Anqing 246011, People's Republic of China}

\date{\today}

\begin{abstract}
We investigate the nucleation of Ising model on complex networks and
focus on the role played by the heterogeneity of degree distribution
on nucleation rate. Using Monte Carlo simulation combined with
forward flux sampling, we find that for a weak external field the
nucleation rate decreases monotonically as degree heterogeneity
increases. Interestingly, for a relatively strong external field the
nucleation rate exhibits a nonmonotonic dependence on degree
heterogeneity, in which there exists a maximal nucleation rate at an
intermediate level of degree heterogeneity. Furthermore, we develop
a heterogeneous mean-field theory for evaluating the free-energy
barrier of nucleation. The theoretical estimations are qualitatively
consistent with the simulation results. Our study suggests that
degree heterogeneity plays a nontrivial role in the dynamics of
phase transition in networked Ising systems.
\end{abstract}
\pacs{89.75.Hc, 64.60.Q-, 05.50.+q} \maketitle

\section{Introduction}\label{sec1}
Since many social, biological, and physical systems can be properly
described by complex networks, dynamics on complex networks have
received considerable attention in recent decades
\cite{RMP02000047,SIR03000167,PRP06000175,PRP08000093}. In
particular, phase transitions on complex networks have been a
subject of intense research in the field of statistical physics and
many other disciplines \cite{RMP08001275}. Owing to the
heterogeneity in degree distribution, phase transitions on complex
networks are drastically different from those on regular lattices in
Euclidean space. For instance, degree heterogeneity can lead to a
vanishing percolation threshold \cite{PhysRevLett.85.4626}, the
whole infection of disease with any small spreading rate
\cite{PRL01003200}, the Ising model to be ordered at all
temperatures \cite{PHA02000260,PLA02000166,PRE02016104}, the
transition from order to disorder in voter models \cite{EPL0768002},
synchronization to be suppressed
\cite{PhysRevLett.91.014101,PhysRevE.71.016116} and different path
towards synchronization in oscillator network
\cite{PhysRevLett.98.034101}, just to list a few. However, there is
much less attention paid to the dynamics of phase transition itself
on complex networks, such as nucleation process in a first-order
phase transition.

Nucleation is a fluctuation-driven process that initiates the decay
of a metastable state into a more stable one \cite{Kashchiev2000}.
Many important phenomena in nature, like crystallization
\cite{Nature011020}, glass formation \cite{PhysRevE.57.5707}, and
protein folding \cite{PNAS9510869} are closely related to nucleation
process. In the context of complex networks, the study of nucleation
process is not only of theoretical importance for understanding how
a first-order phase transition happens in networked systems, but
also may have potential implications in real situations, such as the
transitions between different dynamical attractors in neural
networks \cite{PNAS04004341} and the genetic switch between high and
low-expression states in gene regulatory networks
\cite{PNAS06008372,Plos09004872}, and opinion revolution
\cite{JSM0708026} as well as language replacement
\cite{CCP08000935,ACS0800357} in social networks.

Recently, we have made the first step for studying nucleation
process of Ising model on complex networks, where we have identified
that nucleation pathways using rare-event sampling technique, such
as nucleating from nodes with smaller degree on heterogeneous
networks \cite{PhysRevE.83.031110} and multi-step nucleation process
on modular networks \cite{PhysRevE.83.046124}. In addition, we found
that the size-effect of the nucleation rate on mean-field-type
networks \cite{PhysRevE.83.031110} and nonmonotonic dependence of
the nucleation rate on modularity of networks
\cite{PhysRevE.83.046124}. As mentioned above, degree heterogeneity
has a significant effect on dynamics on complex networks. Therefore,
a natural question arises: how degree heterogeneity affects
nucleation of Ising model on complex networks? To answer this
question, in this paper, we study the dynamics of nucleation on
various network models whose heterogeneity of degree distribution
can be continuously changed by adjusting a single parameter. We use
Monte Carlo (MC) simulation combined with forward flux sampling
(FFS) to compute the nucleation rate and consider the effect of
degree heterogeneity on the rate. Since the critical temperature of
Ising model on uncorrelated random networks increases with the
heterogeneity of degree distribution
\cite{RMP08001275,PHA02000260,PLA02000166,PRE02016104}, one may come
to an intuitive conclusion: if both the temperature and external
field are fixed, the nucleation rate will decrease monotonically as
degree heterogeneity increases. Here, we show that such an intuition
is not the case: the nucleation rate can change monotonically or
nonmonotonically with degree heterogeneity depending on the level of
driving force, i.e., the value of external field. For a weak
external field, the nucleation rate decreases monotonically with
degree heterogeneity, whereas for a relatively strong external field
there exists a maximal nucleation rate corresponding to a moderate
level of degree heterogeneity. Furthermore, we present a
heterogeneous mean-field theory for calculating free-energy barrier
of nucleation. The theoretical results qualitatively agree with the
simulation ones.

\section{Model and Simulation Descriptions}\label{sec2}
The Ising model in a network comprised of $N$ nodes is described by
the Hamiltonian
\begin{equation}
\mathcal {H}=-J\sum\limits_{i < j}{a_{ij}s_i s_j}-h \sum\limits_i
s_i, \label{eq1}
\end{equation}
where spin variable $s_i$ at node $i$ takes either $+1$ (up) or $-1$
(down). $J(>0)$ is the coupling constant and $h$ is the external
field imposed on each node. The elements of the adjacency matrix of
the network take $a_{ij}=1$ if nodes $i$ and $j$ are connected and
$a_{ij}=0$ otherwise.

The simulation is performed by standard Metropolis spin-flip
dynamics, in which we attempt to flip each spin once, on average,
during each MC cycle. In each attempt, a randomly chosen spin is
flipped with the probability $\min(1,e^{- \beta \Delta E})$, where
$\beta=1/(k_B T)$ with the Boltzmann constant $k_B$ and the
temperature $T$, and $\Delta E$ is the energy change due to the
flipping process. We set $h>0$ and $T<T_c$, where $T_c$ is the
critical temperature. The initial configuration is prepared with a
metastable state in which $s_i=-1$ for most of the spins. The system
will stay in that state for a significantly long time before
undergoing a nucleating transition to the thermodynamic stable state
with most spins pointing up.

Since nucleation is an activated process that occurs extremely slow,
brute-force simulation is prohibitively expensive. To overcome this
difficulty, we will use a recently developed simulation method, FFS
\cite{PRL05018104}. This method allows us to calculate nucleation
rate and determine the properties of ensemble toward nucleation
pathways. The simulation results below are obtained by averaging
over at least $5$ independent FFS samplings and $10$ different
network realizations.

\section{Results}\label{sec3}
\begin{figure}
\centerline{\includegraphics*[width=0.85\columnwidth]{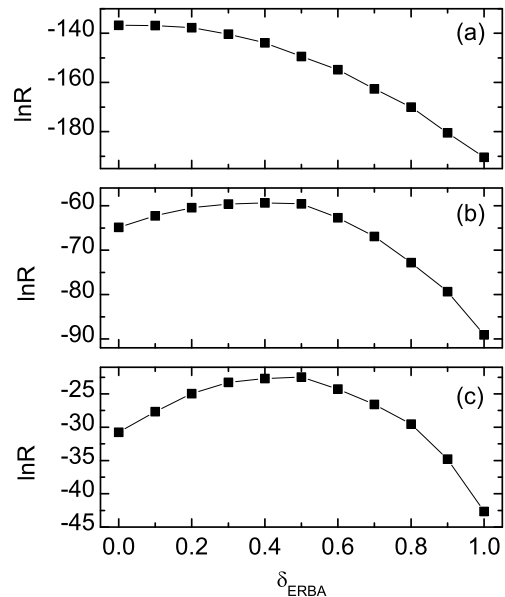}}
\caption{ The logarithm of the nucleation rate $\ln R$ as a function
of the strength of degree heterogeneity $\delta_{ERBA}$ for $h=0.5$
(a), $h=0.8$ (b), and $h=1.0$ (c). Other parameters are $N=1000$,
the mean degree $\left\langle k \right\rangle =6$, and $T=2.5$.
\label{fig1}}
\end{figure}

To study the effect of degree heterogeneity on nucleation, we first
adopt a network model proposed in Ref.\cite{PhysRevE.73.056124}. The
network model allows us to construct networks with the same mean
degree, interpolating from Erdo-Re\`nyi (ER) graphs to
Baraba\`si-Albert (BA) SF networks by tuning a single parameter
$\delta_{ERBA}$. For $\delta_{ERBA}=0$ one gets ER graphs with a
Poissonian degree distribution whereas for $\delta_{ERBA}=1$ the
resulting networks are SF with $P(k) \sim {k^{ - 3}}$. Increasing
$\delta_{ERBA}$ from 0 to 1, the degree heterogeneity of the network
increases. Fig.\ref{fig1} shows that the logarithm of the nucleation
rate $\ln R$ as a function of $\delta$ for three different external
fields: $h=0.5$, 0.8, and 1.0. For $h=0.5$, $\ln R$ decreases
monotonically with $\delta_{ERBA}$, implying that degree
heterogeneity is unfavorable for the occurrence of nucleation
events. Interestingly, for $h=0.8$ $\ln R$ is no longer
monotonically dependent on $\delta_{ERBA}$: as degree heterogeneity
increases, $\ln R$ first increases slowly until $\delta_{ERBA}=0.5$
and then decreases rapidly. Further increasing $h$ to $h=1.0$, $\ln
R$ clearly exhibits a nonmonotonic change with $\delta_{ERBA}$. That
is, there exists a maximal nucleation rate that occurs at a moderate
strength of degree heterogeneity.

\begin{figure}
\centerline{\includegraphics*[width=0.85\columnwidth]{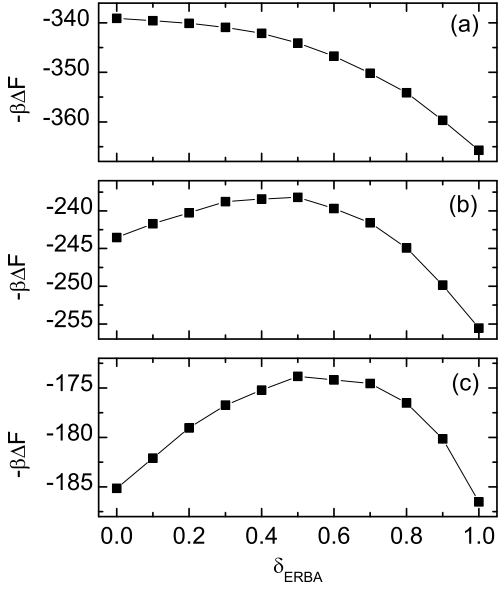}}
\caption{Theoretical results of $-\beta \Delta F$ as a function of
$\delta_{ERBA}$ for $h=0.5$ (a), $h=0.8$ (b), and $h=1.0$ (c). Other
parameters are the same as those in Fig.\ref{fig1}. \label{fig2}}
\end{figure}

To understand the above simulation results, we shall give a
heterogenous mean-field theory on complex networks for evaluating
the nucleation barrier. First, we define $m_k$ as the average
magnetization of a node with degree $k$, i.e., ${m_k} = N_k^{ -
1}\sum\nolimits_{i|{k_i} = k} {{s_i}}$, where $N_k$ is the number of
nodes with degree $k$. Furthermore, for a network without degree
correlation, the probability that a randomly chosen nearest neighbor
node has degree $k$ is ${{kP(k)} \mathord{\left/
 {\vphantom {{kP(k)} {\left\langle k \right\rangle }}} \right.
 \kern-\nulldelimiterspace} {\left\langle k \right\rangle }}$, where $P(k)=N_k/N$ is degree
distribution and $\left\langle k \right\rangle  = \sum\nolimits_k
{kP(k)}$ is the mean degree. Thus, the interaction energy between a
node with degree $k$ and its neighboring nodes can be expressed as
$-J k{m_k}\sum\nolimits_{k'} {k'P(k')} {{{m_{k'}}} \mathord{\left/
{\vphantom {{{m_{k'}}} {\left\langle k \right\rangle }}} \right.
\kern-\nulldelimiterspace} {\left\langle k \right\rangle }}$. The
total energy of the network can be written as
\begin{eqnarray}
  E &=&  - \frac{1}{2}J\sum\limits_k {{N_k}} k{m_k}\sum\limits_{k'} {\frac{{k'P(k'){m_{k'}}}}{{\left\langle k \right\rangle }}}-h\sum\limits_k N_k m_k \nonumber  \hfill
  \\
  & =&  - \frac{1}{2}NJ\left\langle k \right\rangle {{m'}^2}-Nhm,
  \hfill \label{eq2}
\end{eqnarray}
where
\begin{equation}
m' = \sum\limits_k {\frac{{kP(k){m_k}}}{{\left\langle k
\right\rangle }}} \label{eq3}
\end{equation}
is the average magnetization of a randomly chosen nearest neighbor
node, and $m = \sum\nolimits_k {P(k)} {m_k}$ is the average
magnetization of a randomly chosen node. Note that $m'$ differs from
$m$ in general. Special cases for which $m'=m$ are provided by
$k$-independent quantities $m_k=m$. In particular, for the all-spin
down configuration with $m_k=-1$ for all $k$ and for the all-spin-up
configuration with $m_k=1$ for all $k$, one has $m'=m=-1$ and
$m'=m=1$, respectively.

Defining $S_k$ as the entropy of a node with degree $k$, the total
entropy of the network is
\begin{equation}
S = \sum\limits_k {{N_k}} {S_k}=N\sum\limits_k {{P(k)}} {S_k},
\label{eq4}
\end{equation}
with
\begin{equation}
{S_k} =  - {k_B}\left[ {\frac{{1 + {m_k}}}{2}\ln \left( {\frac{{1 +
{m_k}}}{2}} \right) + \frac{{1 - {m_k}}}{2}\ln \left( {\frac{{1 -
{m_k}}}{2}} \right)} \right]. \label{eq5}
\end{equation}
Combining Eq.\ref{eq2} and Eq.\ref{eq4}, we can get the expression
of free energy, i.e., $F=E-T S$.

At the minimum and maximum points of free energy, we have
${{\partial F} \mathord{\left/ {\vphantom {{\partial F} {\partial
{m_k}}}} \right. \kern-\nulldelimiterspace} {\partial {m_k}}} = 0$,
which yields the mean-field equation of $m_k$
\cite{RMP08001275,PLA02000166},
\begin{equation}
{m_k} = \tanh \left[ {\beta h + \beta Jkm'} \right]. \label{eq6}
\end{equation}
Substituting Eq.\ref{eq6} into Eq.\ref{eq3}, we get
\begin{equation}
m' = \sum\limits_k {\frac{{kP(k)}}{{\left\langle k \right\rangle
}}\tanh \left[ {\beta h + \beta Jkm'} \right]}. \label{eq7}
\end{equation}
Eq.\ref{eq7} is a self-consistent equation of $m'$ that can be
numerically solved. In the present settings, Eq.\ref{eq7} has three
solutions: $m'_{-}$, $m'_{0}$, and $m'_{+}$, where $m'_{\pm}$ are
stable solutions and $m'_{0}$ is unstable one. Inserting the three
solutions of $m'$ into the rsh of Eq.\ref{eq6}, we can obtain $m_k$,
and then get $E_{\alpha}$, $S_{\alpha}$ and $F_{\alpha}$
($\alpha={-, 0, +}$) according to Eq.\ref{eq2} and Eq.\ref{eq4}.
Since $h>0$, we have $F_0>F_{-}>F_{+}$, which gives the free-energy
barrier from metastable to stable state $\Delta F=F_0-F_{-}$ and
thus estimate the nucleation rate $R\sim \exp(- \beta \Delta F)$.

Theoretical results of $-\beta \Delta F$ as a function of $\delta$
are shown in Fig.2, where the parameters are the same as those in
Fig.\ref{fig1}. It is clear that the theoretical results are
qualitatively consistent with the simulation ones.

\begin{figure}
\centerline{\includegraphics*[width=1.1\columnwidth]{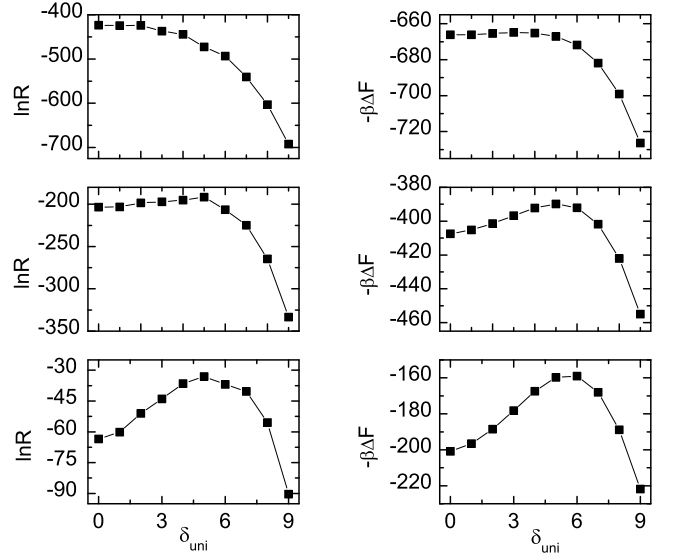}}
\caption{Simulation (left panels) and theoretical (right panels)
results on networks with uniform degree distribution. The external
fields from top to bottom are $h=1.0$, $2.0$, and $3.0$,
respectively. Other parameters are $N=1000$, the mean degree
$\left\langle k \right\rangle =10$, and $T=3$. \label{fig3}}
\end{figure}

\begin{figure}
\centerline{\includegraphics*[width=1.1\columnwidth]{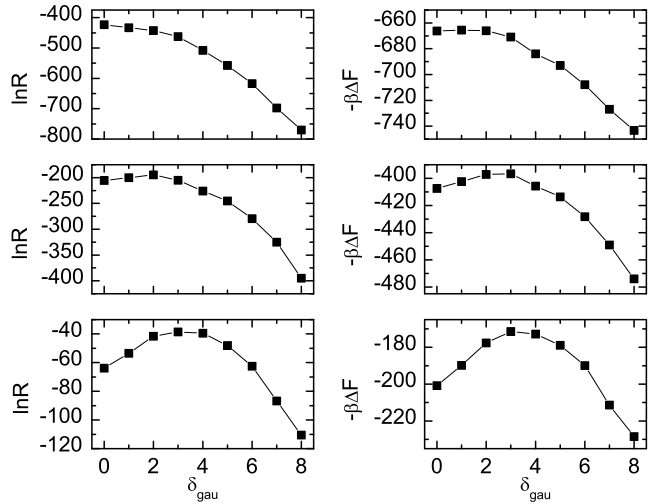}}
\caption{Simulation (left panels) and theoretical (right panels)
results on networks with Gaussian degree distribution. The external
fields from top to bottom are $h=1.0$, $2.0$, and $3.0$,
respectively. Other parameters are $N=1000$, the mean degree
$\left\langle k \right\rangle =10$, and $T=3$. \label{fig4}}
\end{figure}

In order to check the generality of the above results, we shall
calculate nucleation rate on some other network models by both
numerical simulations and theory. Firstly, we construct a network
with uniform degree distribution in which node degree is randomly
selected in the range $[\left\langle k \right\rangle  -
\delta_{uni}, \left\langle k \right\rangle + \delta_{uni}]$, where
$\delta_{uni}$ is an integer between 0 and $\left\langle k
\right\rangle -1$ that controls the strength of degree
heterogeneity. The network is generated according to the Molloy-Reed
algorithm \cite{RSA95161}. This construction eliminates the degree
correlations between neighboring nodes. Fig.\ref{fig3} shows the
simulation and theoretical results, from which the similar phenomena
are also present: for weak external field the nucleation rate
decreases monotonically with degree heterogeneity, while for strong
external field the nucleation rate varies nonmonotonically with
degree heterogeneity. Moreover, we construct a network with Gaussian
degree distribution with fixed mean degree $\left\langle k
\right\rangle$ and variance $\delta_{gau}$. As shown in
Fig.\ref{fig4}, both the simulation and theoretical results display
the similar phenomena.

\section{Summary}\label{sec4}
In summary, using Ising model on complex networks we have shown how
degree heterogeneity affects the rate of nucleation. The main
results of the present paper are that for a weak external field the
nucleation rate decreases monotonically as degree heterogeneity
increases, whereas for a relatively strong external field the
nucleation rate first increases and then decreases with the
increment of degree heterogeneity. Therefore, the nucleation rate
can change monotonically or nonmonotonically with degree
heterogeneity depending on the value of the external field. The
results are robust to different network models, thereby verifying
the generality of the results. Moreover, we have developed the
so-called heterogeneous mean-field theory for calculating the
free-energy barrier to nucleate and thus estimating the nucleation
rate. The theory is effective in qualitatively predicting the
simulation results. Our findings indicate that degree heterogeneity
plays a nontrivial role in the nucleation events of Ising model on
complex networks.

\begin{acknowledgments}
We acknowledge supports from the National Science Foundation of
China (11205002, 20933006, 20873130, and 11147163), ``211 project"
of Anhui University (02303319), and the Key Scientific Research Fund
of Anhui Provincial Education Department (KJ2012A189).
\end{acknowledgments}

% Create the reference section using BibTeX:
%
%\bibliographystyle{apsrev}
%\bibliography{Nucleation_RN}

\begin{thebibliography}{29}
\expandafter\ifx\csname
natexlab\endcsname\relax\def\natexlab#1{#1}\fi
\expandafter\ifx\csname bibnamefont\endcsname\relax
  \def\bibnamefont#1{#1}\fi
\expandafter\ifx\csname bibfnamefont\endcsname\relax
  \def\bibfnamefont#1{#1}\fi
\expandafter\ifx\csname citenamefont\endcsname\relax
  \def\citenamefont#1{#1}\fi
\expandafter\ifx\csname url\endcsname\relax
  \def\url#1{\texttt{#1}}\fi
\expandafter\ifx\csname urlprefix\endcsname\relax\def\urlprefix{URL
}\fi \providecommand{\bibinfo}[2]{#2}
\providecommand{\eprint}[2][]{\url{#2}}

\bibitem[{\citenamefont{Albert and Barab\'{a}si}(2002)}]{RMP02000047}
\bibinfo{author}{\bibfnamefont{R.}~\bibnamefont{Albert}} \bibnamefont{and}
  \bibinfo{author}{\bibfnamefont{A.-L.} \bibnamefont{Barab\'{a}si}},
  \bibinfo{journal}{Rev. Mod. Phys.} \textbf{\bibinfo{volume}{74}},
  \bibinfo{pages}{47} (\bibinfo{year}{2002}).

\bibitem[{\citenamefont{Newman}(2003)}]{SIR03000167}
\bibinfo{author}{\bibfnamefont{M.~E.~J.} \bibnamefont{Newman}},
  \bibinfo{journal}{SIAM Review} \textbf{\bibinfo{volume}{45}},
  \bibinfo{pages}{167} (\bibinfo{year}{2003}).

\bibitem[{\citenamefont{Boccaletti et~al.}(2006)\citenamefont{Boccaletti,
  Latora, Moreno, Chavez, and Hwang}}]{PRP06000175}
\bibinfo{author}{\bibfnamefont{S.}~\bibnamefont{Boccaletti}},
  \bibinfo{author}{\bibfnamefont{V.}~\bibnamefont{Latora}},
  \bibinfo{author}{\bibfnamefont{Y.}~\bibnamefont{Moreno}},
  \bibinfo{author}{\bibfnamefont{M.}~\bibnamefont{Chavez}}, \bibnamefont{and}
  \bibinfo{author}{\bibfnamefont{D.-U.} \bibnamefont{Hwang}},
  \bibinfo{journal}{Phys. Rep.} \textbf{\bibinfo{volume}{424}},
  \bibinfo{pages}{175} (\bibinfo{year}{2006}).

\bibitem[{\citenamefont{Arenas et~al.}(2008)\citenamefont{Arenas,
  D\'iaz-Guilera, Kurths, Moreno, and Zhou}}]{PRP08000093}
\bibinfo{author}{\bibfnamefont{A.}~\bibnamefont{Arenas}},
  \bibinfo{author}{\bibfnamefont{A.}~\bibnamefont{D\'iaz-Guilera}},
  \bibinfo{author}{\bibfnamefont{J.}~\bibnamefont{Kurths}},
  \bibinfo{author}{\bibfnamefont{Y.}~\bibnamefont{Moreno}}, \bibnamefont{and}
  \bibinfo{author}{\bibfnamefont{C.}~\bibnamefont{Zhou}},
  \bibinfo{journal}{Phys. Rep.} \textbf{\bibinfo{volume}{469}},
  \bibinfo{pages}{93} (\bibinfo{year}{2008}).

\bibitem[{\citenamefont{Dorogovtsev et~al.}(2008)\citenamefont{Dorogovtsev,
  Goltseve, and Mendes}}]{RMP08001275}
\bibinfo{author}{\bibfnamefont{S.~N.} \bibnamefont{Dorogovtsev}},
  \bibinfo{author}{\bibfnamefont{A.~V.} \bibnamefont{Goltseve}},
  \bibnamefont{and} \bibinfo{author}{\bibfnamefont{J.~F.~F.}
  \bibnamefont{Mendes}}, \bibinfo{journal}{Rev. Mod. Phys.}
  \textbf{\bibinfo{volume}{80}}, \bibinfo{pages}{1275} (\bibinfo{year}{2008}).

\bibitem[{\citenamefont{Cohen et~al.}(2000)\citenamefont{Cohen, Erez, ben
  Avraham, and Havlin}}]{PhysRevLett.85.4626}
\bibinfo{author}{\bibfnamefont{R.}~\bibnamefont{Cohen}},
  \bibinfo{author}{\bibfnamefont{K.}~\bibnamefont{Erez}},
  \bibinfo{author}{\bibfnamefont{D.}~\bibnamefont{ben Avraham}},
  \bibnamefont{and} \bibinfo{author}{\bibfnamefont{S.}~\bibnamefont{Havlin}},
  \bibinfo{journal}{Phys. Rev. Lett.} \textbf{\bibinfo{volume}{85}},
  \bibinfo{pages}{4626} (\bibinfo{year}{2000}).

\bibitem[{\citenamefont{Pastor-Satorras and Vespignani}(2001)}]{PRL01003200}
\bibinfo{author}{\bibfnamefont{R.}~\bibnamefont{Pastor-Satorras}}
  \bibnamefont{and}
  \bibinfo{author}{\bibfnamefont{A.}~\bibnamefont{Vespignani}},
  \bibinfo{journal}{Phys. Rev. Lett.} \textbf{\bibinfo{volume}{86}},
  \bibinfo{pages}{3200} (\bibinfo{year}{2001}).

\bibitem[{\citenamefont{Aleksiejuk et~al.}(2002)\citenamefont{Aleksiejuk,
  Holysta, and Stauffer}}]{PHA02000260}
\bibinfo{author}{\bibfnamefont{A.}~\bibnamefont{Aleksiejuk}},
  \bibinfo{author}{\bibfnamefont{J.~A.} \bibnamefont{Holysta}},
  \bibnamefont{and} \bibinfo{author}{\bibfnamefont{D.}~\bibnamefont{Stauffer}},
  \bibinfo{journal}{Physica A} \textbf{\bibinfo{volume}{310}},
  \bibinfo{pages}{260} (\bibinfo{year}{2002}).

\bibitem[{\citenamefont{Bianconi}(2002)}]{PLA02000166}
\bibinfo{author}{\bibfnamefont{G.}~\bibnamefont{Bianconi}},
  \bibinfo{journal}{Phys. Lett. A} \textbf{\bibinfo{volume}{303}},
  \bibinfo{pages}{166} (\bibinfo{year}{2002}).

\bibitem[{\citenamefont{Dorogovtsev et~al.}(2002)\citenamefont{Dorogovtsev,
  Goltsev, and Mendes}}]{PRE02016104}
\bibinfo{author}{\bibfnamefont{S.~N.} \bibnamefont{Dorogovtsev}},
  \bibinfo{author}{\bibfnamefont{A.~V.} \bibnamefont{Goltsev}},
  \bibnamefont{and} \bibinfo{author}{\bibfnamefont{J.~F.~F.}
  \bibnamefont{Mendes}}, \bibinfo{journal}{Phys. Rev. E}
  \textbf{\bibinfo{volume}{66}}, \bibinfo{pages}{016104}
  (\bibinfo{year}{2002}).

\bibitem[{\citenamefont{Lambiotte}(2007)}]{EPL0768002}
\bibinfo{author}{\bibfnamefont{R.}~\bibnamefont{Lambiotte}},
  \bibinfo{journal}{Europhys. Lett} \textbf{\bibinfo{volume}{78}},
  \bibinfo{pages}{68002} (\bibinfo{year}{2007}).

\bibitem[{\citenamefont{Nishikawa et~al.}(2003)\citenamefont{Nishikawa, Motter,
  Lai, and Hoppensteadt}}]{PhysRevLett.91.014101}
\bibinfo{author}{\bibfnamefont{T.}~\bibnamefont{Nishikawa}},
  \bibinfo{author}{\bibfnamefont{A.~E.} \bibnamefont{Motter}},
  \bibinfo{author}{\bibfnamefont{Y.-C.} \bibnamefont{Lai}}, \bibnamefont{and}
  \bibinfo{author}{\bibfnamefont{F.~C.} \bibnamefont{Hoppensteadt}},
  \bibinfo{journal}{Phys. Rev. Lett.} \textbf{\bibinfo{volume}{91}},
  \bibinfo{pages}{014101} (\bibinfo{year}{2003}).

\bibitem[{\citenamefont{Motter et~al.}(2005)\citenamefont{Motter, Zhou, and
  Kurths}}]{PhysRevE.71.016116}
\bibinfo{author}{\bibfnamefont{A.~E.} \bibnamefont{Motter}},
  \bibinfo{author}{\bibfnamefont{C.}~\bibnamefont{Zhou}}, \bibnamefont{and}
  \bibinfo{author}{\bibfnamefont{J.}~\bibnamefont{Kurths}},
  \bibinfo{journal}{Phys. Rev. E} \textbf{\bibinfo{volume}{71}},
  \bibinfo{pages}{016116} (\bibinfo{year}{2005}).

\bibitem[{\citenamefont{G\'omez-Garde\~nes
  et~al.}(2007)\citenamefont{G\'omez-Garde\~nes, Moreno, and
  Arenas}}]{PhysRevLett.98.034101}
\bibinfo{author}{\bibfnamefont{J.}~\bibnamefont{G\'omez-Garde\~nes}},
  \bibinfo{author}{\bibfnamefont{Y.}~\bibnamefont{Moreno}}, \bibnamefont{and}
  \bibinfo{author}{\bibfnamefont{A.}~\bibnamefont{Arenas}},
  \bibinfo{journal}{Phys. Rev. Lett.} \textbf{\bibinfo{volume}{98}},
  \bibinfo{pages}{034101} (\bibinfo{year}{2007}).

\bibitem[{\citenamefont{Kashchiev}(2000)}]{Kashchiev2000}
\bibinfo{author}{\bibfnamefont{D.}~\bibnamefont{Kashchiev}},
  \emph{\bibinfo{title}{Nucleation: basic theory with applications}}
  (\bibinfo{publisher}{Butterworths-Heinemann}, \bibinfo{address}{Oxford},
  \bibinfo{year}{2000}).

\bibitem[{\citenamefont{Auer and Frenkel}(2001)}]{Nature011020}
\bibinfo{author}{\bibfnamefont{S.}~\bibnamefont{Auer}} \bibnamefont{and}
  \bibinfo{author}{\bibfnamefont{D.}~\bibnamefont{Frenkel}},
  \bibinfo{journal}{Nature} \textbf{\bibinfo{volume}{409}},
  \bibinfo{pages}{1020} (\bibinfo{year}{2001}).

\bibitem[{\citenamefont{Johnson et~al.}(1998)\citenamefont{Johnson, Mel'cuk,
  Gould, Klein, and Mountain}}]{PhysRevE.57.5707}
\bibinfo{author}{\bibfnamefont{G.}~\bibnamefont{Johnson}},
  \bibinfo{author}{\bibfnamefont{A.~I.} \bibnamefont{Mel'cuk}},
  \bibinfo{author}{\bibfnamefont{H.}~\bibnamefont{Gould}},
  \bibinfo{author}{\bibfnamefont{W.}~\bibnamefont{Klein}}, \bibnamefont{and}
  \bibinfo{author}{\bibfnamefont{R.~D.} \bibnamefont{Mountain}},
  \bibinfo{journal}{Phys. Rev. E} \textbf{\bibinfo{volume}{57}},
  \bibinfo{pages}{5707} (\bibinfo{year}{1998}).

\bibitem[{\citenamefont{Fersht}(1995)}]{PNAS9510869}
\bibinfo{author}{\bibfnamefont{A.~R.} \bibnamefont{Fersht}},
  \bibinfo{journal}{Proc. Natl. Acad. Sci. USA} \textbf{\bibinfo{volume}{92}},
  \bibinfo{pages}{10869} (\bibinfo{year}{1995}).

\bibitem[{\citenamefont{Bar-Yam and Epstein}(2004)}]{PNAS04004341}
\bibinfo{author}{\bibfnamefont{Y.}~\bibnamefont{Bar-Yam}} \bibnamefont{and}
  \bibinfo{author}{\bibfnamefont{I.~R.} \bibnamefont{Epstein}},
  \bibinfo{journal}{Proc. Natl. Acad. Sci. USA} \textbf{\bibinfo{volume}{101}},
  \bibinfo{pages}{4341} (\bibinfo{year}{2004}).

\bibitem[{\citenamefont{Tian and Burrage}(2006)}]{PNAS06008372}
\bibinfo{author}{\bibfnamefont{T.}~\bibnamefont{Tian}} \bibnamefont{and}
  \bibinfo{author}{\bibfnamefont{K.}~\bibnamefont{Burrage}},
  \bibinfo{journal}{Proc. Natl. Acad. Sci. USA} \textbf{\bibinfo{volume}{103}},
  \bibinfo{pages}{8372} (\bibinfo{year}{2006}).

\bibitem[{\citenamefont{Koseska et~al.}(2009)\citenamefont{Koseska, Zaikin,
  Kurths, and Garc\'ia-Ojalvo}}]{Plos09004872}
\bibinfo{author}{\bibfnamefont{A.}~\bibnamefont{Koseska}},
  \bibinfo{author}{\bibfnamefont{A.}~\bibnamefont{Zaikin}},
  \bibinfo{author}{\bibfnamefont{J.}~\bibnamefont{Kurths}}, \bibnamefont{and}
  \bibinfo{author}{\bibfnamefont{J.}~\bibnamefont{Garc\'ia-Ojalvo}},
  \bibinfo{journal}{PLoS ONE} \textbf{\bibinfo{volume}{4}},
  \bibinfo{pages}{e4872} (\bibinfo{year}{2009}).

\bibitem[{\citenamefont{Lambiotte and Ausloos}(2007)}]{JSM0708026}
\bibinfo{author}{\bibfnamefont{R.}~\bibnamefont{Lambiotte}} \bibnamefont{and}
  \bibinfo{author}{\bibfnamefont{M.}~\bibnamefont{Ausloos}},
  \bibinfo{journal}{J. Stat. Mech.} p. \bibinfo{pages}{P08026}
  (\bibinfo{year}{2007}).

\bibitem[{\citenamefont{Ke et~al.}(2008)\citenamefont{Ke, Gong, and
  Wang}}]{CCP08000935}
\bibinfo{author}{\bibfnamefont{J.}~\bibnamefont{Ke}},
  \bibinfo{author}{\bibfnamefont{T.}~\bibnamefont{Gong}}, \bibnamefont{and}
  \bibinfo{author}{\bibfnamefont{W.~S.-Y.} \bibnamefont{Wang}},
  \bibinfo{journal}{Commun. Comput. Phys.} \textbf{\bibinfo{volume}{3}},
  \bibinfo{pages}{935} (\bibinfo{year}{2008}).

\bibitem[{\citenamefont{Wichmann et~al.}(2008)\citenamefont{Wichmann, Stauffer,
  Schulze, and Holman}}]{ACS0800357}
\bibinfo{author}{\bibfnamefont{S.}~\bibnamefont{Wichmann}},
  \bibinfo{author}{\bibfnamefont{D.}~\bibnamefont{Stauffer}},
  \bibinfo{author}{\bibfnamefont{C.}~\bibnamefont{Schulze}}, \bibnamefont{and}
  \bibinfo{author}{\bibfnamefont{E.~W.} \bibnamefont{Holman}},
  \bibinfo{journal}{Adv. Complex Syst.} \textbf{\bibinfo{volume}{11}},
  \bibinfo{pages}{357} (\bibinfo{year}{2008}).

\bibitem[{\citenamefont{Chen et~al.}(2011)\citenamefont{Chen, Shen, Hou, and
  Xin}}]{PhysRevE.83.031110}
\bibinfo{author}{\bibfnamefont{H.}~\bibnamefont{Chen}},
  \bibinfo{author}{\bibfnamefont{C.}~\bibnamefont{Shen}},
  \bibinfo{author}{\bibfnamefont{Z.}~\bibnamefont{Hou}}, \bibnamefont{and}
  \bibinfo{author}{\bibfnamefont{H.}~\bibnamefont{Xin}},
  \bibinfo{journal}{Phys. Rev. E} \textbf{\bibinfo{volume}{83}},
  \bibinfo{pages}{031110} (\bibinfo{year}{2011}).

\bibitem[{\citenamefont{Chen and Hou}(2011)}]{PhysRevE.83.046124}
\bibinfo{author}{\bibfnamefont{H.}~\bibnamefont{Chen}} \bibnamefont{and}
  \bibinfo{author}{\bibfnamefont{Z.}~\bibnamefont{Hou}},
  \bibinfo{journal}{Phys. Rev. E} \textbf{\bibinfo{volume}{83}},
  \bibinfo{pages}{046124} (\bibinfo{year}{2011}).

\bibitem[{\citenamefont{Allen et~al.}(2005)\citenamefont{Allen, Warren, and ten
  Wolde}}]{PRL05018104}
\bibinfo{author}{\bibfnamefont{R.~J.} \bibnamefont{Allen}},
  \bibinfo{author}{\bibfnamefont{P.~B.} \bibnamefont{Warren}},
  \bibnamefont{and} \bibinfo{author}{\bibfnamefont{P.~R.} \bibnamefont{ten
  Wolde}}, \bibinfo{journal}{Phy. Rev. Lett.} \textbf{\bibinfo{volume}{94}},
  \bibinfo{pages}{018104} (\bibinfo{year}{2005}).

\bibitem[{\citenamefont{G\'omez-Garde\~nes and
  Moreno}(2006)}]{PhysRevE.73.056124}
\bibinfo{author}{\bibfnamefont{J.}~\bibnamefont{G\'omez-Garde\~nes}}
  \bibnamefont{and} \bibinfo{author}{\bibfnamefont{Y.}~\bibnamefont{Moreno}},
  \bibinfo{journal}{Phys. Rev. E} \textbf{\bibinfo{volume}{73}},
  \bibinfo{pages}{056124} (\bibinfo{year}{2006}).

\bibitem[{\citenamefont{Molloy and Reed}(1995)}]{RSA95161}
\bibinfo{author}{\bibfnamefont{M.}~\bibnamefont{Molloy}} \bibnamefont{and}
  \bibinfo{author}{\bibfnamefont{B.}~\bibnamefont{Reed}},
  \bibinfo{journal}{Random Struct. Algorithms} \textbf{\bibinfo{volume}{6}},
  \bibinfo{pages}{161} (\bibinfo{year}{1995}).

\end{thebibliography}

\end{document}